\documentclass{mn2e}
\usepackage{epsfig}

\title[The Little Homunculus]{Doppler tomography of the Little Homunculus:
High resolution spectra of [Fe~{\sc ii}] $\lambda$16435 around
Eta Carinae\thanks{Based on observations obtained at the Gemini
Observatory, which is operated by AURA, under a cooperative agreement
with the NSF on behalf of the Gemini partnership: the National Science
Foundation (US), the Particle Physics and Astronomy Research Council
(UK), the National Research Council (Canada), CONICYT (Chile), the
Australian Research Council (Australia), CNPq (Brazil), and CONICET
(Argentina).}}

\author[N.\ Smith]{Nathan Smith\thanks{Email:
nathans@casa.colorado.edu}\thanks{Hubble Fellow} \\ Center for
Astrophysics and Space Astronomy, University of Colorado, 389 UCB,
Boulder, CO 80309, USA}

\date{Accepted 0000, Received 0000, in original form 0000}
\pagerange{\pageref{firstpage}--\pageref{lastpage}} \pubyear{2002}
\def\arcdeg{\degr}
\begin{document}
\label{firstpage}\maketitle

\begin{abstract}

High-resolution spectra of [Fe~{\sc ii}] $\lambda$16435 around $\eta$
Carinae provide powerful diagnostics of the geometry and kinematics of
the ``Little Homunculus'' (LH) growing inside the larger Homunculus
nebula.  The LH expansion is not perfectly homologous: while
low-latitudes are consistent with linear expansion since 1910, the
polar caps imply ejection dates around 1920--1930.  However, the
expansion speed of the LH is much slower than the post-eruption wind,
so the star's powerful wind may accelerate the LH.  With an initial
ejection speed of 200 km s$^{-1}$ in 1890, the LH would have been
accelerated to its present speed if the mass is roughly
0.1~$M_{\odot}$.  This agrees with an independent estimate of the LH
mass based on its density and volume.  In any case, an ejection after
1930 is ruled out.  Using the LH as a probe of the 1890 event, then,
it is evident that its most basic physical parameters (total mass and
kinetic energy; 0.1~$M_{\odot}$ and 10$^{46.9}$ ergs, respectively)
are orders of magnitude less than during the giant eruption in the
1840s.  Thus, the ultimate energy sources were different for these two
events -- yet their ejecta have the same bipolar geometry.  This clue
may point toward a collimation mechanism separate from the underlying
causes of the outbursts.

\end{abstract}


\section{INTRODUCTION}

The Homunculus nebula around $\eta$ Carinae is a key object for
understanding bipolar mass-loss in the late stages of stellar
evolution.  The $\ga$10 M$_{\odot}$ it contains (Smith et al.\ 2003b)
constitutes the major product of $\eta$ Car's giant eruption in the
1840s (Currie et al.\ 1996; Smith \& Gehrz 1998; Morse et al.\ 2001).
However, the process that focussed the prolate mass loss remains
uncertain, even though several ideas have been pursued (e.g., Frank et
al.\ 1995, 1998; Garcia-Segura et al.\ 1997; Owocki \& Gayley 1997;
Dwarkadis \& Balick 1998; Langer et al.\ 1999; Maeder \& Desjacques
2001; Smith 2002b; Smith et al.\ 2003a; Gonzalez et al.\ 2004; Soker
2004; Matt \& Balick 2004).  In this regard, the geometry of
subsequent ejecta inside the Homunculus may provide critical clues.

Ishibashi et al.\ (2003) discovered a smaller nebula called the
``Little Homunculus'', revealed by Doppler shifts of narrow lines in
spectra of $\eta$ Car.  This smaller homuncule gestating inside the
larger one defies the putative gender neutrality of the Homunculus,
but offers an important clue to the mechanism that caused its bipolar
shape.  Ishibashi et al.\ measured proper motions of the Little
Homunculus (LH hereafter), indicating an age of roughly 100 years.
The observed expansion of the LH (much slower than the Homunculus) is
similar to absorption velocities seen in historical spectra from 1893
(Walborn \& Liller 1977; Whitney 1952).  {\it Thus, the LH is most
likely the product of a separate event, but shares the same prolate
geometry as the larger Homunculus.  Whatever the cause of $\eta$ Car's
bipolarity may be, it is persistent.}  There are other hints of
recurring outflow geometry (Smith et al.\ 2004b), including the
present-day bipolar wind (Smith et al.\ 2003a).  Other ejecta have
been attributed to $\eta$ Car's 1890 outburst as well (Smith \& Gehrz
1998; Davidson et al.\ 2001; Smith et al.\ 2004b), adding weight to
the LH's putative origin in that event (but see Dorland et al.\ 2004).

The LH is not seen in direct visual-wavelength images because it is
overwhelmed by starlight scattered off dust in the Homunculus, while
emission from the LH is also obscured by that dust.  However, some
emission structures inside the Homunculus can be recognized by their
temporal variability or wavelength dependence (Smith et al.\ 2000,
2004a, 2004b); the ``Purple Haze'' and emission knots seen in [S~{\sc
iii}] and [N~{\sc ii}] may be parts of the LH.

Fortunately, the LH exhibits very bright emission from infrared (IR)
lines of [Fe~{\sc ii}], most notably [Fe~{\sc ii}] $\lambda$16435
(Smith 2002b).  This line can be enhanced in shocks or
photodissociation regions, and bright [Fe~{\sc ii}] $\lambda$16435 is
common in nebulae of other luminous blue variables (Smith 2002a).  The
high-resolution spectra of [Fe~{\sc ii}] $\lambda$16435 presented here
significantly advance our understanding of the kinematics of the LH.
They have higher dispersion and better sensitivity than earlier data,
and the infrared [Fe~{\sc ii}] line can more easily penetrate the
intervening dust screen.

Although thermal-IR images do not show the LH, presumably due to
insufficient dust, they have revealed a bright dust torus, marking the
point in the equator where the two lobes of the Homunculus meet (Smith
et al.\ 2002, 2003b).  Since this dust torus and the LH both occupy
similar projected areas on the sky (within $\sim$2$\arcsec$ of the
star), the relationship between them is ambiguous.  Using [Fe~{\sc
ii}] $\lambda$16435 emission to compare the spatial extents of the LH
and dust torus is one goal of this paper, in addition to constraining
other properties of the LH and the 1890 outburst.


\section{OBSERVATIONS}

High-resolution ($R\simeq$60,000, $\sim$5 km s$^{-1}$) near-IR spectra
of $\eta$ Car were obtained on 2004 May 15 using the Phoenix
spectrograph on the Gemini South telescope (Hinkle et al.\ 2003).
Phoenix has a 1024$\times$256 InSb detector with a pixel scale of
0$\farcs$085$\times$1.4 km s$^{-1}$ at a wavelength of $\sim$1.6
$\mu$m.  Sky conditions were photometric, and the seeing was
0$\farcs$3--0$\farcs$4.  Removal of airglow lines was accomplished by
subtracting an observation of an off-source position 35$\arcsec$
southeast.

The 0$\farcs$34-wide long-slit aperture was oriented at
P.A.=310$\arcdeg$ along the polar axis of the Homunculus (Fig.\ 1).
To sample the kinematics across the LH, the slit was positioned on the
bright central star, plus offsets of 1$\arcsec$ and 2$\arcsec$ in
either direction perpendicular to the slit axis as shown in Fig.\ 1.
At each slit position, three pairs of 60-second exposures sampled
[Fe~{\sc ii}] 16435, for a total on-source exposure time in middle
regions of the Homunculus of 6 min.

HR~5571 was observed with Phoenix on the same night with the same
grating setting in order to correct for telluric absorption.  Telluric
lines were used for wavelength calibration, using the telluric
spectrum available from NOAO.  Velocities were calculated adopting a
vacuum rest wavelength of 16439.981 \AA \ for the [Fe~{\sc ii}]
$\lambda$16435 ($a^4F-a^4D$) line, and these velocities were corrected
to a heliocentric reference frame.  (Heliocentric velocities will be
quoted here.)  Uncertainty in the resulting velocities is $\pm$1 km
s$^{-1}$, dominated by scatter in the dispersion solution for telluric
lines.

Figure 2 shows the resulting long-slit data for [Fe~{\sc ii}] at the
five different slit positions, where the bright reflected continuum
light in the Homunculus has been subtracted out to enhance the
contrast of the line emission.  Structures near the center of each
panel (at $\pm$250 km s$^{-1}$) are from the LH, while filaments from
the Homunculus are seen at larger velocities (see Smith 2002b).  The
vertical dashed lines in Fig.\ 1 mark the systemic velocity of $\eta$
Car at --8.1 km s$^{-1}$, measured from earlier Phoenix spectra of
H$_2$ in the Homunculus (Smith 2004).  The H$_2$ systemic velocity is
more reliable than that given by narrow H$\alpha$ emission (Boumis et
al.\ 1998), because of asymmetric and variable ionization structure in
ejecta near the star (e.g., Smith et al.\ 2004a, 2004b).


\begin{figure}\begin{center}
\epsfig{file=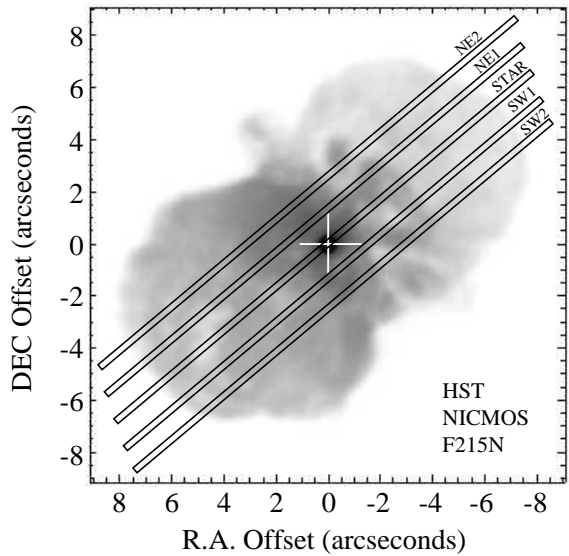,width=3in}\end{center}
\caption{Phoenix slit aperture positions superposed on a 2~$\micron$
HST/NICMOS image of $\eta$ Car from Smith \& Gehrz (2000).}
\end{figure}

\begin{figure}\begin{center}
\epsfig{file=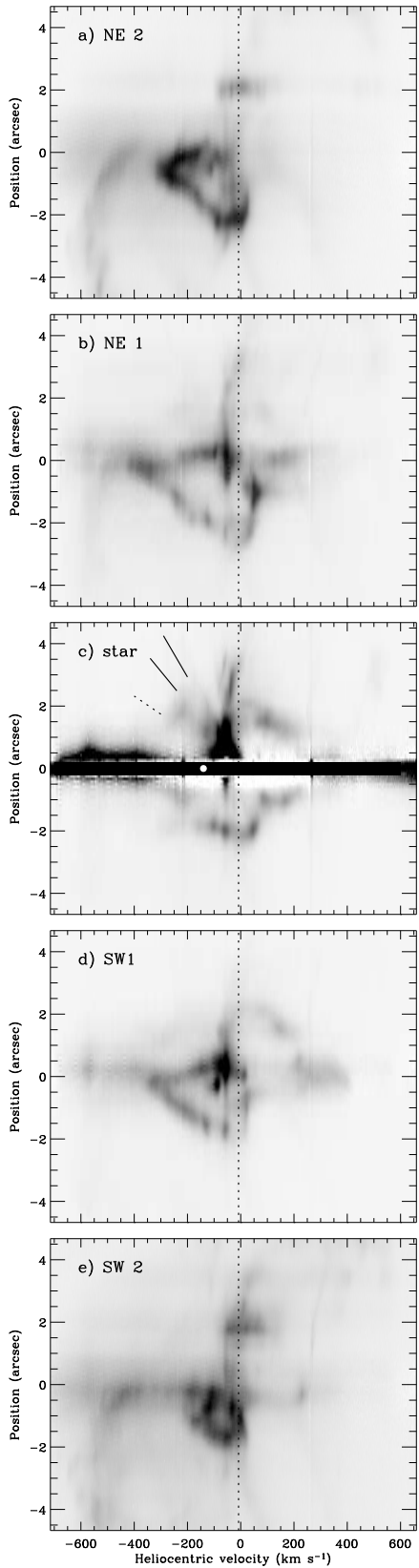,height=8.9in}\end{center}
\caption{(a) Long-slit kinematics of [Fe~{\sc ii}] $\lambda$16435
around $\eta$ Car at the five slit positions shown in Fig.\ 1.
$V_{\rm sys}$=$-$8.1 km s$^{-1}$ (Smith 2004) is marked with a dashed
line.}
\end{figure}


\section{KINEMATIC STRUCTURE}

With a few exceptions due to extinction (see below), all five slit
positions in Figure 2 confirm the basic bipolar structure inferred by
Ishibashi et al.\ (2003).  In general, the LH is a miniature version
of the larger Homunculus, although the general shape appears more
scrunched in the polar direction. In every case the polar cap on the
blueshifted lobe is clearly seen, while it was harder to discern in
earlier data at visual wavelengths (Ishibashi et al.\ 2003).  Thus,
the star's much faster stellar wind has not yet cleared a path through
the slower LH; such resistance by the LH may have important
consequences for its mass, momentum, and acceleration (see \S 4 and \S
6).  However, in general, [Fe~{\sc ii}] emission in Figure 2 gives the
impression that the LH is clumpy, with a filling factor of perhaps
0.5. The five slit placements sampled differences in the
kinematic structure, as described below:

{\it NE2 (Fig.\ 2a)}: Of the five slit positions shown in Figure 2,
NE2 displays the most asymmetric structure, because the redshifted/NW
polar lobe of the LH is essentially invisible.  This implies that the
NE2 slit position encounters more extinction in the equator than at
other positions, blocking the light from the far side of the nebula.
Indeed, this region located 2$\arcsec$--3$\arcsec$ north of the star
is relatively dark in optical images and has a high optical depth of
cool dust (Smith et al.\ 2003b).  The only sign of the NW lobe is a
faint blotch at 0 km s$^{-1}$, +2$\arcsec$, which coincides with a
feature in the equatorial skirt at visual wavelengths.  This signifies
that bright portions of the skirt may be holes where light can
penetrate.  The SE polar lobe is bright, and has a morphology
consistent with a slice through a flattened polar bubble.

{\it NE1 (Fig.\ 2b)}: The kinematic structure at this position already
differs from NE2.  Part of the redshifted NW lobe can be seen,
although it still suffers more extinction than the blueshifted lobe.
Here the SE lobe has a more angular or trapezoidal shape than at NE2,
with straight side walls, pointed corners, and a flat polar cap.
Velocities as fast as --400 km s$^{-1}$ are seen.  An interesting
feature at the NE1 position is the pair of bright spots that occupy
the equator of the LH, giving the impression of a slice through a
tilted equatorial ring (also seen in channel maps of [Fe~{\sc ii}]
$\lambda$4891 presented by Ishibashi et al., and in gasdynamical
simulations of the LH; Gonzalez et al.\ 2004).  These are the
kinematic counterparts of the Purple Haze and [S~{\sc iii}] and
[N~{\sc ii}] features in images, which show marked temporal
variability (Smith et al.\ 2000; 2004a).  At the display scale chosen
here (see \S 4), a line drawn through these two features would trace
out an equatorial plane tilted from the plane of the sky by
$\sim$40$\arcdeg$, consistent with the inclination of the Homunculus
(Smith 2002b; Davidson et al.\ 2001).

\begin{figure*}\begin{center}
\epsfig{file=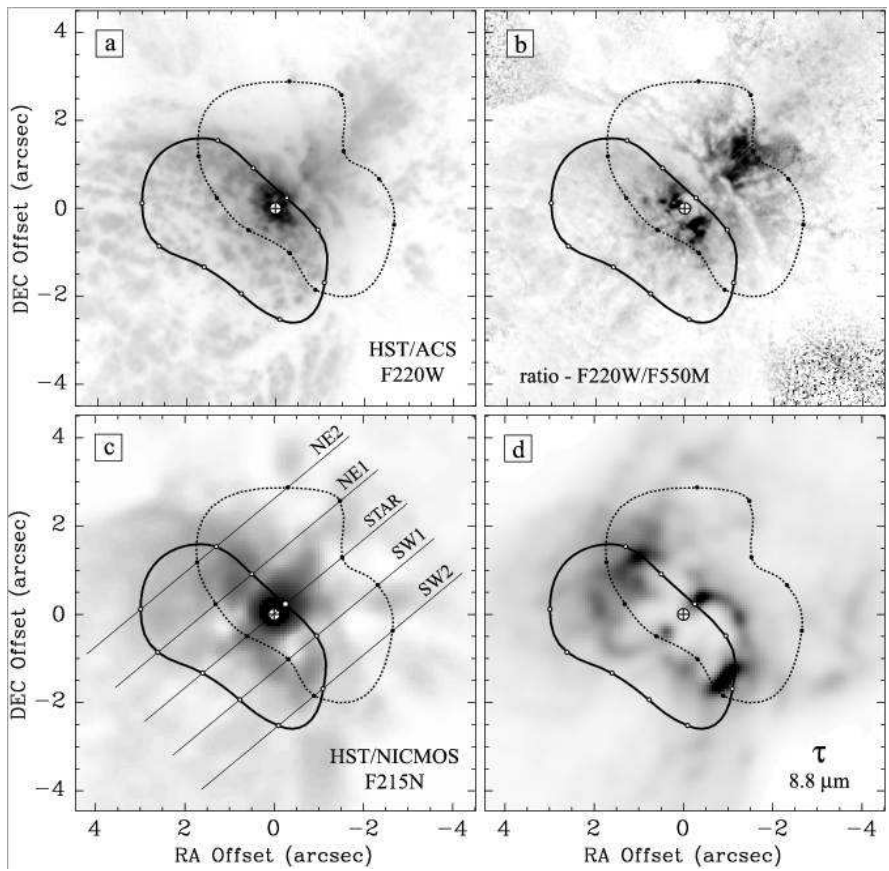,width=4.6in}\end{center}
\caption{The spatial extents of the blueshifted (solid) and redshifted
(dotted) polar lobes of the LH superposed on various images of $\eta$
Car: (a) 2200~\AA\ image (Smith et al.\ 2004b); (b) 2200 \AA/5500 \AA\
flux ratio image showing the UV excess emission from the ``Purple
Haze'' (Smith et al.\ 2004b); (c) near-IR 2.15 $\mu$m continuum image
(Smith \& Gehrz 2000); (d) 8.8 $\mu$m optical depth or the column
density of warm dust (Smith et al.\ 2003b).  Filled and unfilled
circles mark the measured extremities of the LH lobes along each of
the five slit positions, while the curves are interpolated between
these points.}
\end{figure*}

{\it STAR (Fig.\ 2c)}: Both polar lobes of the LH are clearly seen at
this position, indicating that here we can see to the far side of the
LH (Smith et al.\ 1998; Smith \& Gehrz 2000; Davidson et al.\ 2001;
Smith 2002b).  Interestingly, this is also the only slit position
exhibiting blueshifted equatorial [Fe~{\sc ii}] emission from the
``skirt'' of the Homunculus.\footnote{This position also shows
blueshifted He~{\sc i} $\lambda$10830 (Smith 2002b), [Ni~{\sc ii}]
$\lambda$7379 (Davidson et al.\ 2001), [Sr~{\sc ii}] (Hartman et al.\
2004), and narrow H91$\alpha$ (Duncan et al.\ 1997).  At least two
different velocity components are seen here -- a slow narrow component
and a somewhat broader fast component at a different tilt angle (see
also Hartman et al.\ 2004; Smith 2002b; Davidson et al.\ 2001).  The
two solid diagonal lines in Figure 2$c$ are the tilt angles of the
equator in these plots for ejection dates of 1843 and 1890, and the
dashed line is for 1940.  The narrow component lies close to the
expected tilt for 1843, while the broader equatorial component seems
to have been ejected after 1890 but well before 1940 (see Smith \&
Gehrz 1998; Dorland et al.\ 2004; Smith et al.\ 2004b).}  The
saturated stellar continuum makes it impossible to measure the precise
velocity of the LH along the line of sight, but considering the
kinematic structure in adjacent slit positions, the blueshifted wall
of the LH crosses our line of sight to the star at --140$\pm$20 km
s$^{-1}$ (shown with the white dot).  This agrees with the UV
absorption component at --146 km s$^{-1}$ (Gull et al.\ 2004),
confirming that the absorption feature is indeed from the LH.  The
bright [Fe~{\sc ii}] emission just northwest of the star at --46 km
s$^{-1}$ (Smith 2004) is from the Weigelt knots (Weigelt \&
Ebersberger 1986).

{\it SW1 (Fig.\ 2d)}: The SW1 position gives the best representation
of the bipolar structure of the LH, with two nearly complete, closed
polar lobes.  It exhibits the same flat-topped or trapezoidal
structure as is seen at NE1.  However, SW1 begins to show asymmetry in
the LH; the redshifted NW lobe is about 20\% larger than the SE lobe.
Also, unlike NE1, there is a strong brightness asymmetry in the
emission knots associated with the equatorial ring (the blueshifted
knot is much brighter, as for the slit passing through the
star).

{\it SW2 (Fig.\ 2e)}: Structure at this position further accentuates
the asymmetry of the LH.  The blueshifted lobe is much smaller than
its counterpart at NE2, and it is about half the size of its own
redshifted lobe.  Obscuration of the redshifted lobe by dust in the
equator is apparent.

\section{THE AGE OF THE LITTLE HOMUNCULUS}

The shape of the LH and the way its kinematic structure is displayed
in Figure 2 can give valuable clues to its age.  The relative scale
between the spatial and velocity directions in Figure 2 reflects the
age, in the sense that a horizontal stretch implies younger material,
and horizontal compression implies older material.  Since the LH has
some inherent asymmetry, the choice of the horizontal stretch is
subjective, depending on which features one uses to gauge the
appropriate scaling.  Figure 2 is displayed with a scale of
1$\arcsec$=117 km s$^{-1}$, corresponding to an ejection date around
1910.  This was chosen to match proper motions of the Weigelt knots
(Smith et al.\ 2004a).  Also, at this scale, a line drawn through the
two bright equatorial knots at the NE1 position is tilted from
vertical by $\sim$40$\arcdeg$, which matches the inclination of the
Homunculus, as noted above.

However, at this display scale, some portions of the LH still appear
somewhat stretched, which has two ramifications.  First, if one
assumes axial symmetry, it confirms that the Homunculus was not
ejected during the Great Eruption in the 1840s, because such features
would be horizontally compressed, rather than elongated.  Second, the
polar features look the most symmetric at a display scale
corresponding to a later ejection date of 1920--1930.

The potential reconciliation of the age discrepancy may have to do
with acceleration of ejecta, as suggested already by Smith et al.\
(2004b).  Material ejected in the 1890 event that has been accelerated
by radiation pressure or stellar wind ram pressure would show faster
Doppler shifts and higher proper motion than expected.  As noted
already, equatorial zones of the LH have probably been accelerated
since an 1890 ejection (Smith et al.\ 2004b).  In Figure 2 the polar
features appear elongated even for an ejection date of 1910 --- if
they originated in the 1890 event as well, they must have been
accelerated even more than the corresponding equatorial features.  In
polar directions, ram pressure of the wind probably dominates, since
$\eta$ Car has a latitude-dependent wind with an effective mass-loss
rate of $\dot{M}\simeq$10$^{-3} M_{\odot}$ yr$^{-1}$ and polar speeds
of 600 km s$^{-1}$ (Smith et al.\ 2003a).  Thus, the polar wind speed
is {\it much} faster than the LH, so we should expect some
interaction.  The fact that the polar caps of the LH have so far
maintained their integrity means that the fast stellar wind has not
been able to plow through the LH or disrupt it through Rayleigh-Taylor
instabilitites, so momentum is being transferred from the stellar wind
to the LH.  The acceleration depends on the mass of the LH, of course,
which is investigated below in \S 6.  In any case, the stretch of the
polar caps indicates that the expansion of the LH is not perfectly
homologous like its larger counterpart.

\begin{figure}\begin{center}
\epsfig{file=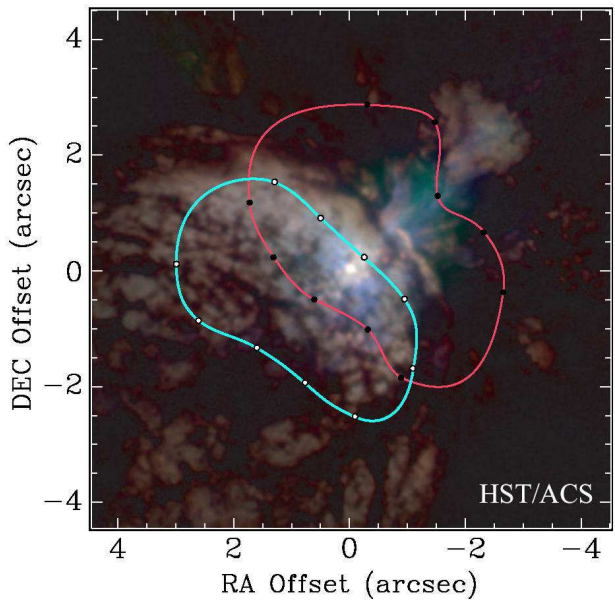,width=3.2in}\end{center}
\caption{The spatial extents of the blueshifted (blue) and redshifted
(red) polar lobes of the LH superposed on a 3-color {\it HST}/ACS
image of $\eta$ Car from Smith et al.\ (2004b)}
\end{figure}

\section{SPATIAL EXTENT}

In lieu of a direct image of the LH, it is useful to investigate its
projected appearance using spatial information gleaned from Figure 2.
The measured extremeties of the blueshifted and redshifted polar lobes
along each of the five slits are shown in Figures 3 and 4, superposed
on various imaging data from previous studies.  Obviously, some
artistic license was taken in drawing these smooth curves, especially
beyond the NE2 and SW2 slit positions where no additional information
is available.  Nevertheless, these curves give a fair depiction of the
overall extent of the LH.

The LH has no outstanding correspondence with any of the clumps and
filaments seen in scattered light in normal UV or visual-wavelength
images of $\eta$ Car, although it does match the spatial extent of the
``Purple Haze'' (Figs.\ 3$a$ and 4).  This correlation is most
striking in Figure 3$b$, where the Purple Haze emission is almost
entirely within the boundaries of the LH.  The brightest UV excess
within 1$\arcsec$ to the NE and SW of the star is in the overlap
region of the two polar lobes of the LH, where one expects to find the
LH's equatorial features.  At near-IR wavelengths where one sees
through the dust in the SE polar lobe of the Homunculus, there is no
correlation between the LH and scattered near-IR continuum light
(Figure 3$c$).

It is quite evident from Figure 3$d$ that the LH and the disrupted
``dust torus'' seen in the thermal-IR (Smith et al.\ 2002) are not the
same physical structure, although there may be an interesting
relationship between them.  The NW edge of the blueshifted LH lobe and
the SE edge of the redshifted lobe both seem to hug the inside edges
of the dust torus.  This implies that the dust torus is really an
equatorial ring, and would impede expansion of the LH at low
latitudes, while the LH is free to expand out the poles.  The distinct
lack of dust in the interior regions supports the conjecture that the
bright [Fe~{\sc ii}] emission in the LH arises because Fe atoms are
not locked up in grains.

Interestingly, the best spatial corresponce between the LH and
features seen in images is with variable radio continuum structures
(Duncan et al.\ 1997).  The brightest radio continuum features are
presumably equatorial, but the low-level contours of 3 cm emission
coincide spatially with the extent of the LH.  Furthermore, while the
NW equatorial feature shows narrow blueshifted H91$\alpha$ emission,
the fainter surrounding emission shows broad lines with widths of
$\pm$250 km s$^{-1}$ (Duncan et al.\ 1997).  This velocity range
agrees well with the kinematics of the LH.

\section{MASS AND KINETIC ENERGY}

\subsection{Mass Estimate \# 1:  Density and Geometry}

One way to estimate the mass of the LH is to simply deduce values for
the volume and average density.  Figure 2 indicates that the polar
caps of the LH can be approximated as two disks, each with a radius of
1$\farcs$6 (5.9$\times$10$^{16}$ cm) and thickness 0$\farcs$4
(1.4$\times$10$^{16}$ cm).  Thus, the volume occupied by the polar
caps of the LH is roughly 3$\times$10$^{50}$ cm$^3$ (this includes
both poles).  Approximating the side walls of the LH as a pair of
truncated funnels with the same thickness gives roughly the same
volume as contributed by the caps, for a total volume of
$V\simeq$6$\times$10$^{50}$ cm$^3$.  Then the mass of the LH is given
by

\begin{equation}
M_{LH} = \mu m_H \frac{n_e}{\chi}\,f\,V
\end{equation}

\noindent where $\mu\simeq$1.25 for a He mass fraction $Y$=0.4, $\chi$
is the hydrogen ionization fraction, and $f$ is a filling factor of
$\sim$0.5.

Near-IR [Fe~{\sc ii}] line ratios suggest an electron density of
roughly 10$^{4.2}$ cm$^{-3}$ in the LH (Smith 2002b).  However, given
the presence of molecular hydrogen (Smith 2002b) and weakness of
H$\alpha$ in the Homunculus, the ionization fraction in the LH is
probably low, although not as low as in the more distant lobes of the
Homunculus.  Calculations with the {\sc cloudy} spectral synthesis
code intended to explain the ionization structure of the Homunculus
give $\chi\simeq$10$^{-3}$ at the inner edge of its [Fe~{\sc ii}] zone
(G.\ Ferland, private comm.). Much closer to the star in the Weigelt
knots, $\chi$ rises to 0.5 (Verner et al.\ 2002).  Thus, $\chi$=0.05
is probably a reasonable intermediate value to choose for the LH,
where the radiation field is more intense than in the Homunculus.
With $n_e$=10$^{4.2}$, $f$=0.5, and $\chi$=0.05, the total mass of the
LH would be $\sim$0.1~$M_{\odot}$.

\subsection{Mass Estimate \# 2:  Inertia}

Another way to deduce the mass of the LH is to ask how much inertia is
needed to avoid excessive acceleration by the stellar wind over an
assumed age $t$.  For example, if the LH were too ``light'', $\eta$
Car's powerful stellar wind would quickly accelerate the LH up to
nearly 600 km s$^{-1}$, which would violate the LH's observed current
expansion speed of $\sim$250 km s$^{-1}$ (Fig.\ 2).  The horizontal
elongation of the polar regions of the LH in Figure 2 implies that the
poles have been accelerated more than equatorial zones.  This may
result from higher momentum flux in the polar wind (Smith et al.\
2003a).

Under the assumption of acceleration by the ram pressure of a steady
stellar wind $\rho v^2$, where $v=v_{\infty}-v$ is the relative
velocity between $\eta$ Car's polar wind and the changing speed of the
LH, $v=v(t)$, the equation of motion gives

\begin{equation}
\int\frac{dv}{(v_{\infty}-v)^2} = \frac{\dot{M} t}{m v_{\infty}}
\end{equation}

\noindent where $\dot{M}$ is $\eta$ Car's average mass loss rate
during this time, and $m$ is the mass of a spherical shell accelerated
by the wind.  This can be integrated to yield a relation for the mass
of the LH in terms of the initial ejection velocity of the LH, $u$,
and the current observed difference between the stellar wind and the
speed of the LH, $\Delta v = v_{\infty}-v_{LH}$, given by

\begin{equation}
M_{LH} = \frac{\Omega}{4\pi}\,f\,\dot{M}\,t\,
 \Big{[}\frac{1-u/v_{\infty}}{v_{\infty}/\Delta v - u/\Delta v - 1}\Big{]}
\end{equation}

\noindent where $\Omega$ is the solid angle of the LH as seen by the
star, and $f$ is a filling factor or efficiency factor for the
momentum transfer (i.e. $f\times\Omega/4\pi$ corrects for the fact
that the LH is not a uniform spherical shell with mass $m$).  From the
observed geometry of the LH in Figure 2, $\Omega\simeq$2$\times$2
ster.  Furthermore, from observations we can adopt $\dot{M}$=10$^{-3}
M_{\odot}$ yr$^{-3}$, $v_{\infty}$=600 km s$^{-1}$ (e.g., Hillier et
al.\ 2001), and a present value for the polar speed of the LH of
$v_{LH}$=250 km s$^{-1}$, so that $\Delta v = v_{\infty}-v_{LH}$ = 350
km s$^{-1}$.  Then, the remaining quantities are $t$, $u$, and
$M_{LH}$.

In the most plausible scenario, where the LH was ejected in the 1890
event, we have $t$=114 yr.  Furthermore, spectra obtained in 1893
showed absorption features at about --200 km s$^{-1}$ (Walborn \&
Liller 1977; Whitney 1952), which gives a plausible value for $u$.
With these constraints, equation (3) gives $M_{LH} = f\times0.17 \,
M_{\odot}$.  The similar values in methods 1 and 2 are somewhat
misleading, since method 2 only applies to the polar region (about 1/2
of the mass).  In any case, these arguments suggest that the likely
mass of the LH is of order 0.1--0.2 $M_{\odot}$.  This is consistent
with the lower mass estimated by Ishibashi et al.\ (2003).

This rough agreement between the two independent mass estimates is
encouraging, and adds additional support for an 1890 ejection and
subsequent acceleration of the LH by the 20th-century stellar wind.
If the LH was ejected after 1890 or if the acceleration has not been
constant, then the problem obviously becomes more difficult.  The
alternative scenario -- where the LH and the Weigelt knots were
instead ejected in the early 20th century (Smith et al.\ 2004b;
Dorland et al.\ 2004; Ishibashi et al.\ 2003) -- seems less likely.
Higher masses are required if the LH was ejected after 1890, because
it would need to more effectively resist acceleration.  For example,
for an ejection date of 1930 and only 1\% acceleration, the required
mass is about 1~$M_{\odot}$ (with the same filling factor $f$=0.5).
Masses much above 0.1 $M_{\odot}$ begin to seem highly implausible,
since the LH is invisible in images and has not formed large
quantities of dust.  In other words, {\it the present day expansion
speed of the LH rules out an ejection date as late as 1940}.


\subsection{Kinetic Energy and the 1890 Outburst}

With a mass of 0.1--0.2 $M_{\odot}$ in the LH, Doppler shifts allow
one to estimate the kinetic energy released in the 1890 event.  Most
of the kinetic energy will be contained in the $\sim$0.1 $M_{\odot}$
of material in the two polar caps of the LH, moving at 250 km
s$^{-1}$, while a small fraction is contributed by mass in the side
walls moving at slower speeds.  Then, the total kinetic energy
released in the 1890 event is about 10$^{46.9}$ ergs.

This is a factor of 500 less kinetic energy than was expelled in the
Great Eruption (Smith et al.\ 2003b), signifying that the ultimate
sources of energy for the 1890 and 1840's events were very different.
Furthermore, averaged over the 7 yr duration of the 1890 outburst
(Humphreys et al.\ 1999), the mass-loss rate was
$\dot{M}\simeq$0.02~$M_{\odot}$.  This was only $\sim$4\% of that
during the Great Eruption, and the momentum imparted to the 1890
ejecta was only about 1--2\% of that during the larger event.
Finally, in the 1890 event the ratio of mechanical-to-radiative
luminosity was about 0.02, compared to numbers closer to unity for the
Great Eruption itself (here I have assumed that the bolometric
radiative luminosity was a constant 5$\times$10$^6$ $L_{\odot}$ during
the 1890 event, whereas it increased by about a factor of 4--5 in the
1840's; Humphreys et al.\ 1999).  Thus, the physics of the mass
ejection was also quite different in the two 19th century outbursts.

It is also instructive to compare the 1890 event with the present-day
stellar wind of $\eta$~Carinae.  The average mass-loss rate of the
1890 event was about 20 times higher than the present-day stellar
wind, while the wind momentum was about 10 times stronger.  Thus, the
1890 event was indeed an outburst, in the sense that the mass loss was
enhanced compared to the normal stellar wind parameters.
Interestingly, however, the ratio of mechanical-to-radiative
luminosity in 1890 was about 2\% -- only slightly more than the value
for the present-day stellar wind -- so the acceleration of the LH used
up a similar fraction of available luminosity.  In a radiatively
driven wind, it is often useful to know the wind's ``performance
number'' (the ratio of wind momentum to photon momentum), given by

\begin{equation}
\zeta = \frac{\dot{M}\,v}{L/c}
\end{equation}

\noindent where $L\simeq$5$\times$10$^6~L_{\odot}$ is the radiative
luminosity during the 1890 event.  Thus, $\zeta_{1890}\simeq$50, while
for the present day stellar wind $\zeta\simeq$5 (note that
$\zeta\ga$10$^3$ during the Great Eruption; Smith et al.\ 2003b).
Performance numbers as high as $\sim$10 are typically expected for
very dense line-driven winds (Lucy \& Abbott 1993; Springmann \& Puls
1998), so if mass-loss during the 1890 outburst was a line-driven
wind, it was certainly pushing the limits of that acceleration
mechanism.

\section{SUMMARY: PERSISTENTLY BIPOLAR}

The main results of this study are the following:

1.  The kinematic structure of the LH would seem to suggest ejection
    dates between 1910 and 1930 if one assumes linear expansion and
    rough axial symmetry.  However, linear expansion may be an invalid
    assumption: the polar caps of the LH appear to be intact,
    suggesting that the faster post-eruption wind has not yet broken
    through the LH and may be accelerating it.  Thus, even though the
    expansion of the LH is non-homologous, it may all have been
    ejected during the 1890 eruption if it has been accelerated by ram
    pressure of the post-eruption wind.  In this case, the polar caps
    of the LH have been accelerated more than low-latitudes.

2.  Various clues indicate a total mass for the LH of roughly 0.1
    $M_{\odot}$, so the kinetic energy released in the 1890 event was
    roughly 10$^{46.9}$ ergs.  Thus, the 1890 event was
    orders-of-magnitude less powerful than the Great Eruption in the
    1840's, indicating that the two events had a different energy
    source and probably a different root cause.

Despite these differences, both eruptions gave rise to similar bipolar
geometry with the same polar axis.  This may point toward an {\it
external} collimation mechanism.  For example, while internal
processes may have brought about $\eta$ Car's phenomenal energy
release and mass ejection during the 1840's and again 50 years later,
something else may have helped to collimate the outflow. $\eta$ Car is
thought to be a close binary system (Damineli et al.\ 2000), so one
can certainly envision a scenario where the two stars interact
violently during close periastron passages.  This is by no means a new
suggestion (e.g., Innes 1914), but difficult 3-D calculations are
needed to proceed beyond mere speculation.

In this regard, however, it is interesting to note that some planetary
nebulae surrounding symbiotic binary stars have nested bipolar nebulae
that remind one of the LH and Homunculus of $\eta$ Car.  Two salient
examples are Hb~12 (Hora et al.\ 2000; Welch et al.\ 1999) and
He~2-104 (Corradi et al.\ 2001).  Hb~12 is particularly interesting in
that the smaller bipolar nebula has [Fe~{\sc ii}] $\lambda$16435
emission, while the larger bipolar shell emits near-IR lines of
molecular hydrogen (Welch et al.\ 1999), just like the LH and
Homunculus around $\eta$ Car (Smith 2002b).  Of course, the fact that
these symbiotic planetary nebulae have also had sporadic outbursts
with the same recurring bipolar geometry does not mean that they share
the same collimation mechanism as $\eta$ Car, but the nebular
similarities are intriguing.

On the other hand, the present-day stellar wind is also bipolar and
shares the same axis as the Homunculus (Smith et al.\ 2003a).  Thus,
some intrinsic mechanism that persistently sends material poleward may
be at work in $\eta$ Car as well (e.g., Owocki \& Gayley 1997; Matt \&
Balick 2004).

\smallskip\smallskip\smallskip\smallskip
\noindent {\bf ACKNOWLEDGMENTS}
\smallskip\scriptsize

\noindent Support was provided by NASA through grant HF-01166.01A from
the Space Telescope Science Institute, which is operated by the
Association of Universities for Research in Astronomy (AURA), Inc.,
under NASA contract NAS 5-26555.  These data were obtained in service
observing mode, and I thank the Gemini staff for their assistance.

\label{lastpage}

\begin{thebibliography}
\scriptsize

\bibitem[]{} Boumis, P., Meaburn, J., Bryce, M., \& Lopez, J.A.\ 1998,
MNRAS, 294, 61

\bibitem[]{} Corradi, R.L.M., Livio, M., Balick, B., Munari, U., \&
Schwarz, H.E.\ 2001, ApJ, 553, 211

\bibitem[]{} Currie, D.G., et al.\ 1996, AJ, 112, 1115

\bibitem[]{} Damineli, A., Kaufer, A., Wolf, B., Stahl, O., Lopes,
D.F., \& de Ara\'{u}jo, F.X.\ 2000, ApJ, 528, L101

\bibitem[]{} Davidson, K., Smith, N., Gull, T.R., Ishibashi, K., \&
Hillier, D.J.\ 2001, AJ, 121, 1569

\bibitem[]{} Dorland, B.N., Currie, D.G., \& Hajian, A.R.\ 2004, AJ,
127, 1052

\bibitem[]{} Duncan, R.A., White, S.M., Reynolds, J.E., \& Lim, J.\
1997, in ASP Conf.\ Ser.\ 179, Eta Carinae at the MIllenium, eds.\
J.A.\ Morse, R.M.\ Humphreys, \& A.\ Damineli (san Francisco: ASP), 54


\bibitem[]{} Frank, A., Balick, B., \& Davidson, K.\ 1995, ApJ, 441, L77

\bibitem[]{} Frank, A., Ryu, D., \& Davidson, K.\ 1998, ApJ, 500, 291

\bibitem[]{} Garcia-Segura, G., Langer, N., \& MacLow, M.M.\ 1997, in
ASP Conf.\ Ser.\ 120, Luminous Blue Variables: Massive Stars in
Trabsition, ed.\ A. Nota \& H.J.G.L.M.\ Lamers (san Francisco: ASP),
332

\bibitem[]{} Gonzalez, R., de Gouveia Dal Pino, E.M., Raga, A.C., \&
Velazquez, P.F.\ 2004, ApJ, 600, L59

\bibitem[]{} Gull, T.R., et al.\ 2004, submitted

\bibitem[]{} Hartman, H., Gull, T.R., Johansson, S., Smith, N., \& the
HST Eta Car Treasury Project Team 2004, A\&A, 419, 215

\bibitem[]{} Hillier, D.J., Davidson, K., Ishibashi, K., \& Gull,
T.R.\ 2001, ApJ, 553, 837

\bibitem[]{} Hinkle, K.H., Blum, R., Joyce, R.R., Ridgeway, S.T.,
Rodgers, B., Sharp, N., Smith, V., Valenti, J., \& van der Bliek, N.\
2003, Proc.\ SPIE 4834, 353

\bibitem[]{} Hora, J.L., et al.\ 2000, in ASP Conf.\ Ser.\ 199,
Asymmetrical Planetary Nebulae II: From Origins to Microstructures,
eds.\ J.H. Kastner, N.\ Soker, \& S.\ Rappaport (San Francisco: ASP),
267

\bibitem[]{} Humphreys, R.M., Davidson, K., \& Smith, N.\ 1999, PASP,
111, 1124

\bibitem[]{} Innes, R.T.A.\ 1914, MNRAS, 74, 697

\bibitem[]{} Ishibashi, K., et al.\ 2003, AJ, 125, 3222

\bibitem[]{} Langer, N., Garcia-Segura, G., \& MacLow, M.M.\ 1999,
ApJ, 520, L49

\bibitem[]{} Lucy, L.B., \& Abbott, D.C.\ 1993, ApJ, 405, 738

\bibitem[]{} Maeder, A., \& Desjacques, V.\ 2001, A\&A, 372, L9

\bibitem[]{} Matt, S., \& Balick, B.\ 2004, ApJ, 615, 921

\bibitem[]{} Morse, J.A., Kellogg, J.R., Bally, J., Davidson, K.,
Balick, B., \& Ebbets, D.\ 2001, ApJ, 548, L207

\bibitem[]{} Owocki, S.P., \& Gayley, K.G.\ 1997, in ASP Conf.\ Ser.\
120, Luminous Blue Variables: Massive Stars in Trabsition, ed.\
A. Nota \& H.J.G.L.M.\ Lamers (san Francisco: ASP), 121

\bibitem[]{} Smith, N.\ 2002a, MNRAS, 336, L22

\bibitem[]{} Smith, N.\ 2002b, MNRAS, 337, 1252

\bibitem[]{} Smith, N.\ 2004, MNRAS, 351, L15

\bibitem[]{} Smith, N., Davidson, K., Gull, T.R., Ishibashi, K., \&
Hilier, D.J.\ 2003a, ApJ, 586, 432

\bibitem[]{} Smith, N., \& Gehrz, R.D.\ 1998, AJ, 116, 823

\bibitem[]{} Smith, N., \& Gehrz, R.D.\ 2000, ApJ, 529, L99

\bibitem[]{} Smith, N., Gehrz, R.D., \& Krautter, J.\ 1998, AJ, 116, 1332

\bibitem[]{} Smith, N., Gehrz, R.D., Hinz, P.M., Hoffmann, W.F.,
Mamajek, E.E., Meyer, M.R., \& Hora, J.L.\ 2002, ApJ, 567, L77

\bibitem[]{} Smith, N., Gehrz, R.D., Hinz, P.M., Hoffmann, W.F., Hora,
J.L., Mamajek, E.E., \& Meyer, M.R.\ 2003b, AJ, 125, 1458

\bibitem[]{} Smith, N., Morse, J.A., Davidson, K., \& Humphreys, R.M.\
2000, AJ, 120, 920

\bibitem[]{} Smith, N., Morse, J.A., Collins, N.R., \& Gull, T.R.\
2004a, ApJ, 610, L105

\bibitem[]{} Smith, N., et al.\ 2004b, ApJ, 605, 405

\bibitem[]{} Soker, N.\ 2004, ApJ, 612, 1060

\bibitem[]{} Springmann, U., \& Puls, J.\ 1998, in ASP Conf.\ Ser.\
131, Boulder-Munich II: Properties of Hot Luminous Stars, ed.\ I.D.\
Howarth (San Francisco: ASP), 286

\bibitem[]{} Verner, E.M., Gull, T.R., Bruhweiler, F., Johansson, S.,
Ishibashi, K., \& Davidson, K.\ 2002, ApJ, 581, 1154

\bibitem[]{} Walborn, N.R., \& Liller, M.\ 1977, ApJ, 211, 181

\bibitem[]{} Weigelt, G., \& Ebersberger, J.\ 1986, A\&A, 163, L5

\bibitem[]{} Welch, C.A., Frank, A., Pipher, J.L., Forrest, W.J., \&
Woodward, C.E.\ 1999, ApJ, 522, L69

\bibitem[]{} Whitney, C.A.\ 1952, Harvard Obs.\ Bull., 921, 8

\end{thebibliography}
\end{document}